\documentclass[%
 aip,
 jmp,%
 amsmath,amssymb,
 preprint,%
]{revtex4-1}

\usepackage{graphicx}
\usepackage{dcolumn}
\usepackage{bm}
\usepackage{color}
\usepackage{caption}

\begin{document}

\title[Statistical analysis and modeling of intermittent transport events in the tokamak SOL]{Statistical analysis and modeling of intermittent transport events in the tokamak scrape-off layer}

\author{Johan Anderson}
\email{anderson.johan@gmail.com.}
\affiliation{ 
Department of Earth and Space Sciences, Chalmers University of Technology, SE-412 96 G\"{o}teborg, Sweden
}%
\author{Federico D.~Halpern} 
\affiliation{%
\'{E}cole Polytechnique F\'{e}d\'{e}rale de Lausanne (EPFL), Centre de Recherches en Physique des Plasmas, CH-1015 Lausanne, Switzerland
}%
\author{Pavlos Xanthopoulos}%
\affiliation{ 
Max-Planck-Institut f\"{u}r Plasmaphysik, IPP-Euratom Association, \\ Teilinstitut Greifswald, D-17491 Greifswald, Germany
}%
\author{Paolo Ricci} 
\affiliation{%
\'{E}cole Polytechnique F\'{e}d\'{e}rale de Lausanne (EPFL), Centre de Recherches en Physique des Plasmas, CH-1015 Lausanne, Switzerland
}%
\author{Ivo Furno} 
\affiliation{%
\'{E}cole Polytechnique F\'{e}d\'{e}rale de Lausanne (EPFL), Centre de Recherches en Physique des Plasmas, CH-1015 Lausanne, Switzerland
}%

\date{\today}

\begin{abstract}
The turbulence observed in the scrape-off-layer of a tokamak is often characterized by intermittent events of bursty nature, a feature which
raises concerns about the prediction of heat loads on the physical boundaries of the device. It appears thus necessary to delve into the 
statistical properties of turbulent physical fields such as density, electrostatic potential and temperature, focusing on the 
mathematical expression of tails of the probability distribution functions. The method followed here is to
generate statistical information from time-traces of the plasma density stemming from Braginskii-type fluid simulations, and check this against a first-principles theoretical model. The analysis of the numerical simulations indicates that the probability distribution function of the intermittent process contains strong exponential tails,
as predicted by the analytical theory.

\end{abstract}

\maketitle

\section{\label{sec:level1} Introduction}
Turbulence is a fascinating open problem cutting across scientific boundaries. In magnetic fusion research, understanding
turbulence is a key element for the theoretical explanation of the heat and particle transport in tokamak devices both in the core and edge regions, including the much-debated formation of the plasma pedestal in the high-confinement regime. In particular, understanding the turbulent behavior of the plasma in the most external region of a tokamak, the scrape-off-layer (SOL), has important implication for the operation of present and future devices, such as ITER \cite{p1, p2}.

 The  plasma profiles in the SOL region form from the balance between the plasma outflowing from the tokamak core, 
turbulence transport and end losses at the physical boundary (limiter or divertor) of the device.
This turbulent dynamics in the SOL is characterized by large fluctuations with amplitudes comparable to the background plasma,
and can manifest itself in radially-propagating, coherent meso-scale modes called 'blobs', which have been suggested
to carry (together with streamers) a significant fraction of the heat transport through rare avalanche-like events. \cite{p3,p4,a12,a13,a15} Blobs are
typically intermittent events with a patchy spatial and bursty temporal structure and are responsible for deviations of the probability
distribution function (PDF) --  in the form of exponential tails -- from the Gaussian prediction based on the traditional mean-field theory. \cite{p5}
Controlling the edge heat flux loads, which depend on the instant amplitude of fluctuations, as opposed to the mean load, calls for a
thorough understanding of intermittency, both in terms of analytical modelling and numerical investigations.

A pivotal idea to study intermittency has been to associate the bursty event with the creation of a coherent structure. A candidate that could describe the creation process of the structure is the instanton, which is localized in time and lives during the formation of the coherent structure. The instanton method is a non-perturbative way of calculating PDF tails, which was adopted from quantum field theory and then modified to classical statistical physics for Burgers turbulence and a passive scalar model. \cite{a18,a19} For instance, using the instanton method, it has been shown in Reference \onlinecite{p6} that the PDF tails of momentum flux $R$ are significantly enhanced over the Gaussian prediction. More specifically, the tail exhibits a ubiquitous scaling of the form $\exp{(-\xi R^{3/2})}$, where the coefficient $\xi$ contains all the model-dependent information. \cite{p6}    

In this work, we investigate the statistics governing SOL turbulence by employing first-principles numerical simulations and theoretical analysis alike. 
Starting point for both approaches is a drift-reduced set of the Braginskii fluid equations, \cite{p7,p8} which describes interchange driven turbulence. Our scope is
to confirm theoretical predictions about the behavior of the PDF for the density through numerical results stemming from the simulations. 
By modeling the plasma outflowing from the core as a time-independent source, we further exclude from our model the coupling of SOL turbulence with the plasma dynamics inside the LCFS. For instance, it has been shown that far-SOL simulations of typical L-mode turbulence in the inner-wall-limited Alcator C-Mod configuration manifest similar statistical properties when compared with experimental observations using gas-puff imaging \cite{subm}. This points to the possibility that turbulent structures traveling through the SOL are generated near the LCFS.

The layout of the
paper is as follows: In Section \ref{sec1} we present the drift-reduced Braginskii equations, followed by the analytical modeling in Section \ref{sec2},
where a generalized system of stochastic partial differential equations is presented, as an extension of the Braginskii system. In the same section, we derive
the properties of the PDF for the density. Section \ref{sec3} deals with the numerical simulations and the mathematical processing of the output data, in order
to reconcile these with the theoretical results. Finally, we provide a short summary of the work in Section \ref{sec4}.
\section{Drift-reduced model for tokamak SOL turbulence} \label{sec1}
For the present study, we use a cold-ion drift-reduced model, which can be derived from the Braginskii two-fluid equations\cite{p7} by imposing the orderings $d/dt\ll\omega_{ci}$, $k_\bot\gg k_\|$, and $T_i \ll T_e$. Particle trapping is negligible since $\nu_{\star}\gg 1$ in the SOL of limited plasmas, while finite Larmor radius effects are small since $k_\theta\rho_s\sim 0.1$ for the dominant modes in the non-linear stage. Since the plasma is relatively cold, a fluid model can capture the essential physical ingredients of this system. The drift-reduced equations, in normalized units, read as follows~\cite{p8},
\begin{eqnarray}
\frac{\partial n}{\partial t} =&-\frac{1}{B_0 \rho_*}\left[ \phi,n \right] -\nabla_\| \left(nv_{\|e}\right) + \frac{2}{B_0}\left[\hat{C}\left(p_e\right) - n\hat{C}\left(\phi\right) \right] + D_n\nabla_\perp^2 n + S_{n}\label{eq_dense}\\
\frac{\partial \omega}{\partial t} =&-\frac{1}{B_0 \rho_*}\left[ \phi,\omega \right] -v_{\|i}\nabla_\| \omega + \frac{2B_0}{n}\hat{C}\left(p_e\right) + \frac{B_0^2}{n}\nabla_{\|}j_{\|}+ \frac{B_0}{3n}\hat{C}\left(G_i\right)+D_{\omega}\nabla_\perp^2 \omega \label{eq_omega}\\
\frac{\partial v_{\|e}}{\partial t} =&-\frac{1}{B_0 \rho_*}\left[\phi,v_{\|e}\right]-v_{\|e}\nabla_\| v_{\|e}+D_{v_{\|e}}\nabla_\perp^2 v_{\|e}\nonumber\\
&+\frac{m_i}{m_e}\left( \nu\frac{j_{\|}}{n} + \nabla_\|\phi -\frac{1}{n}\nabla_\| p_e - 0.71\nabla_\| T_e -\frac{2}{3n}\nabla_\| G_e \right)\label{eq_ohm}\\
\frac{\partial v_{\|i}}{\partial t}=&-\frac{1}{B_0 \rho_*}\left[\phi,v_{\|i}\right]-v_{\|i}\nabla_{\|}v_{\|i}-\frac{1}{n}\nabla_{\|}p_e -\frac{2}{3n}\nabla_\| G_i+D_{v_{\|i}}\nabla_\perp^2 v_{\|i}\label{eq_vpari}\\
\frac{\partial T_e}{\partial t}=&-\frac{1}{B_0 \rho_*}\left[\phi,T_{e}\right] -v_{\|e}\nabla_{\|}T_e +\frac{4}{3}\frac{T_e}{B_0}\left[\frac{7}{2}\hat{C}\left(T_e\right) +\frac{T_e}{n}\hat{C}\left(n\right)-\hat{C}\left(\phi\right)\right] \nonumber\\
&+\frac{2}{3}T_e\left(\frac{0.71}{n}\nabla_{\|}j_{\|} - \nabla_{\|}v_{\|e}\right) + D_{T_{e}}\nabla_\perp^2 T_e + S_{T_e}\label{eq_tempe}.
\end{eqnarray}

The equations are given in dimensionless form, with the following normalizations being used: $t=\tilde{t}/(\tilde{R}/\bar{c}_s)$, $\nabla_\bot=\bar{\rho}_s\tilde{\nabla}_\bot$, $\nabla_\|= \tilde{R}\tilde{\nabla}_\|$, $v_\|=\tilde{v_\|}/\bar{c}_s$, $n=\tilde{n}/\bar{n}$, $T_{e}=\tilde{T}_e/\bar{T}_e$, $\phi=e\tilde{\phi}/\bar{T}_e$, $\rho_* = \bar{\rho_s}/\tilde{R}$, $B_0=\tilde{B}/\bar{B}$. Here, the tildes denote quantities in MKS physical units, and the bars denote reference quantities defined in terms of the normalized density $\bar{n}$, the normalized temperature $\bar{T}_e$, and the reference magnetic field $\bar{B}$. All variables are expressed in their dimensionless form unless specified otherwise.
The parallel current is given by $j_\|=n\left(v_{\|i}-v_{\|e} \right)$, while $\nu=e^2 n \tilde{R}/(\tilde{m_i} \sigma_\| \bar{c}_s)$ is the normalized Spitzer resistivity. The vorticity is defined as $\omega=\nabla^2_\bot\phi$, and Eq. \ref{eq_omega} has been simplified using the Boussinesq approximation $\nabla\cdot\left(n d_t\nabla_\bot\phi \right)\approx n d_t\nabla_\bot^2\phi$. 

In the non-linear simulations, plasma outflow from the closed flux surface region is mimicked using density and temperature sources, $S_n$ and $S_{T_e}$, respectively. The terms $G_e$ and $G_i$ represent the gyroviscous part of the pressure tensor (see Ref. \cite{p8}). Small perpendicular diffusion terms of the form $D_f\nabla_\perp^2 f$ are added in order to damp grid-scale modes arising from numerical discretization. In addition, $\left[f,g\right]={\bf b}_0\cdot\left(\nabla f \times \nabla g\right)$ is the Poisson bracket, while $\hat{C}\left(f\right) = (B_0/2)\left[\nabla\times\left({\bf b}_0/B_0\right)\right]\cdot\nabla f$ is the curvature operator.

We consider a SOL model in circular geometry with a toroidal limiter set at the high field side equatorial midplane. The (right-handed) coordinate system used is $(y,x,\varphi)$, where $x$ is the radial coordinate ($x=0$ at the LCFS), $y=x\theta$ is the poloidal distance, and $\varphi$ is the toroidal angle. Under these assumptions, the curvature operator reduces to $\hat{C}\left(f\right)=(\sin\theta)\partial_x f+(\cos\theta+\hat{s}\theta\sin\theta)\partial_y f$ and the Poisson bracket is defined as $\left[f,g\right]=a^{-1}(\partial_y f \partial_x g - \partial_x f \partial_y g)$ ($\hat{s}=(a+r) q^\prime/q$ is the magnetic shear). 

Finally, the plasma interfaces with the vacuum vessel through a magnetized pre-sheath where the fluid drift approximation breaks down. The validity of the drift-reduced model, therefore, formally extends until the magnetic pre-sheath entrance, where we apply the boundary conditions derived in Ref.~\onlinecite{Loizu2012}.

\section{Statistical model of intermittent events} \label{sec2}
Common features of the PDFs inferred from bursty and intermittent processes are strongly non-Gaussian tails while being unimodal in structure. \cite{a10} There exist several ways to derive such PDFs for a physical process, for instance employing the instanton method (see, e.g., Refs. \cite{justin, p6, kim, anderson1, anderson2, a22, a24}) and using the Fokker-Planck method (see, e.g., Refs. \cite{sancho1982, liang2004, hasegawa2008, kim2}). In order to model the intermittent transport events at the edge, a generalized physical model is adapted from the normalized 3D reduced Braginskii equations (\ref{eq_dense})-(\ref{eq_tempe}), presented in Section \ref{sec1}, to be used in the Fokker-Planck method. To derive the model equations we have, for simplicity, replaced the Ohm's law Eq. (\ref{eq_ohm}) by $\nabla_\parallel \phi = \nu j_\parallel$, also have neglected the parallel couplings in Eqs. (\ref{eq_dense}) and (\ref{eq_tempe}), and the ion parallel velocity, Eq. (\ref{eq_vpari}). Replacing the Ohm's law by a simple resistive response is the reason of the $C_\parallel$ term in Eq. (\ref{eq:1.1}). Furthermore, the equation for the electron temperature perturbations, Eq. (\ref{eq_tempe}), and the equation of electron density perturbations, Eq. (\ref{eq_dense}), are merged into one equation for the electron pressure $\pi_e$, where $\pi_e = \log p_e = \log T_e + \log n_e$. In addition, we have neglected the thermal force in Braginskii's model. After these modifications, we arrive at the system of equations 
\begin{eqnarray}
\frac{\partial \nabla^2 \phi}{\partial t} & = & - \frac{1}{B_0 \rho_*} [\phi, \nabla^2 \phi] - 2 \hat{C} (\pi_e) + \hat{C}_{||} \phi + f_0, \label{eq:1.1} \\
\frac{\partial \pi_e }{\partial t} & = & - \frac{1}{B_0 \rho_*} [ \phi, \pi_e ] + \frac{2}{3} \hat{C} (\phi) + \frac{10}{3} \hat{C}(\pi_e), \label{eq:1.2}\\
\hat{C} (\chi) & = & \frac{\partial \chi}{\partial y}, \label{eq:1.3}\\
\hat{C}_{||} (\chi) & = & \frac{\partial^2 \chi}{\partial z^2} \label{eq:1.4}\\
\left[ \phi, \chi \right] & = & \frac{\partial \phi}{\partial x} \frac{\partial \chi}{\partial y} - \frac{\partial \phi}{\partial y} \frac{\partial \chi}{\partial x}. \label{eq:1.5}
\end{eqnarray}
Here, $f_0$ is a zero-mean Gaussian white-noise stochastic forcing that has been added to the drift-reduced Braginskii model, $\hat{C}_{||}$ represents the parallel dynamics, and $B_0$ is the magnitude of the equilibrium magnetic field. The model consists of two coupled non-linear equations with a stochastic forcing in the vorticity equation, Eq. (\ref{eq:1.1}). In order to extract the salient features leading to intermittent events, further manipulations are needed. In particular, while we consider a linearized pressure equation, yielding a linear coupling between the pressure and the potential, a multiplicative stochastic term is introduced to make up for the lack of nonlinear coupling,
 \begin{eqnarray} \label{eq:1.6}
\left[ \phi, \pi_e \right] & \approx & - \frac{1}{B_0 \rho_*}\left( -\frac{\partial \pi_e}{\partial x} \frac{\partial \phi}{\partial y}\right)  = \frac{1}{B_0 \rho_*} \frac{\pi_e}{L_p} \frac{\partial \phi}{ \partial y}.
\end{eqnarray}
Here, the factor $\frac{\pi_e}{L_p} $ will be represented by a zero-mean Gaussian stochastic force $f_1$. Notice that for the sake of generality, both Gaussian forces $f_0$ and $f_1$ are retained. The dynamics is now represented by one equation for the potential with an additive and a multiplicative noise term. Since we consider only the statistics of time-traces, we make use of the coherent structures as traveling solutions of the form
\begin{eqnarray} \label{eq:1.7}
\phi(x,y,t) & = &\psi (x, y - Ut) F(t), \\
\pi_e(x,y,t) & = & \alpha(v, f_1) \phi(x,y,t),  
\end{eqnarray}
where the potential $\phi$ and pressure $\pi_e$ follow each other with a relation between the potential and the pressure as
\begin{eqnarray} \label{eq:1.8}
\alpha(v, f_1) = \frac{f_1 + 2/3}{v + 10/3}.
\end{eqnarray}
The traveling solution propagates perpendicular to the density gradient, however the main transport direction is radial. This enables us to reduce the problem to a time dependent problem, where we find the stochastic equation in $F$ of the form,
\begin{eqnarray}\label{eq:1.9}
\nabla^2_{\perp} \frac{\partial F (t)}{\partial t} & = & v \nabla^2_{\perp} \frac{\partial \psi}{\partial y} F(t) - \frac{1}{B_0 \rho_*} \left[\psi, \nabla^2_{\perp} \psi \right] F^2(t) + \frac{4/3}{v + 10/3} \frac{\partial \psi}{\partial y} F(t)  \nonumber \\
& - & \frac{k_{||}^2}{\nu} \psi F(t) + \frac{2}{v+ 10/3} f_1 \frac{\partial \psi}{\partial y} F(t) + f_0.
\end{eqnarray}
Note that all spatial dependent terms will be treated as constants in the time dependent stochastic equation, which once rewritten, it becomes
\begin{eqnarray}
\frac{\partial F}{\partial t} & = & G_0 F - G_1 F^2 + G_2 F f_1 + f_0, \label{eq:1.10} \\
N G_0 & = & v \nabla^2_{\perp} \frac{\partial \psi}{\partial y} + \frac{4/3}{v + 10/3} \frac{\partial \psi}{\partial y} - \frac{k_{||}}{\nu} \psi, \label{eq:1.11}\\ 
N G_1 & = & \frac{1}{B_0 \rho_*}\left[ \psi, \nabla^2_{\perp} \psi\right] ,\label{eq:1.12}\\
N G_2 & = & \frac{2}{v+10/3} \frac{\partial \psi}{\partial y}, \label{eq:1.13}\\
N & = & \nabla^2_{\perp} \psi. \label{eq:1.14}
\end{eqnarray}
The spatial function $\psi$ is approximately determined by the modon solution of the form $\nabla_{\perp}^2 \psi = -k^2 \psi + C x$, where $k$ is the modon number and $C = \frac{1}{B_0 \rho_*} (v - \frac{2}{\alpha})$ with $\alpha = - \frac{2/3}{v + 10/3}$. However the symmetry is broken by the parallel coupling and is only exact for $k_{||} = 0$. The coefficients can be determined by an average over the coherent structure as $\langle \chi \rangle = \int dx dy \chi \psi = \tilde{\chi}$ (compare the results obtained in Refs. \cite{a22,kim}). In this work it is sufficient to tune the constant through statistical analysis performed on the numerical solutions, remembering that these will vary at different radial positions and structures. Stochastic differential equations with multiplicative noise have been studied earlier and this particular class of dynamical equations have a closed analytical solution presented in Refs. \onlinecite{sancho1982, liang2004, hasegawa2008}. The solution depends on the cross-correlations between the additive noise $f_0$ and the multiplicative noise $f_1$ with a rather complicated solution, however assuming that also the correlation is white in time gives the relation,
\begin{eqnarray} \label{eq:1.15}
\langle f_0(t) f_1(t^{\prime})\rangle = D \delta (t - t^{\prime}).
\end{eqnarray}
We will only present a short note on how to obtain the PDF, following Ref. \cite{sancho1982}. In the Stratonovich interpretation we have the Fokker-Planck equation for $P(F,t) = \langle \delta(F(t) - F)\rangle$,
\begin{eqnarray}\label{eq:1.16}
\frac{\partial P}{\partial t} = - \frac{\partial}{\partial F} (\tilde{G}_0(F) + \tilde{G}_1(F)) + \frac{\partial}{\partial F} \tilde{G}_2(F) \frac{\partial}{\partial F} \tilde{G}_2(F) P(F,t).
\end{eqnarray}
The general stable PDF solution to the approximate Fokker-Planck equation is then,
\begin{eqnarray} \label{eq:1.17}
P(F) & = & \frac{N_0}{|F|} e^{\frac{\tilde{G}_0}{\tilde{G}_2^2} \ln |F| - \frac{\tilde{G}_1}{\tilde{G}_2^2} |F|},
\end{eqnarray}
with the white noise assumption. Note that the PDF of the electrostatic potential $\phi$ and the pressure $\pi_e$ will be the same by construction, and that we have predicted a single-moment quantity, such as the potential in terms of the time dependent term $F$, which implies that the statistics for the density fluctuations is the same. [We note in passing, that the statistics of fluxes require additional work, which is omitted here.]
In conclusion, the sought PDF  for the intermittent and bursty physics in the SOL will have the exponential form
\begin{eqnarray} \label{eq:1.18}
P(n_e) & = & \frac{N_0}{|n_e|} e^{\frac{\tilde{G}_0}{\tilde{G}_2^2} \ln |n_e| - \frac{\tilde{G}_1}{\tilde{G}_2^2} |n_e|},
\end{eqnarray}
which heavily relies on the nonlinear terms in the model. If, instead, such terms are neglected, the model only predicts a Gaussian PDF for the physical quantities. 

\section{Numerical simulations of tokamak SOL transport} \label{sec3}
In this section, we present the outcome of nonlinear simulations of SOL turbulence and the statistical analysis thererof. The simulations are performed using the GBS code, a numerical implementation of Eqs.~(\ref{eq_dense})-(\ref{eq_tempe}). Solving this system of equations involves approximating the spatial derivatives using standard second order accurate finite differences in space. The Arakawa scheme \cite{Arakawa1966} has been employed for the Poisson brackets, while the time advance is carried out using a standard fourth order accurate Runge-Kutta method. The code is fully 3D and flux-driven, which avoids the typical flux-tube partial linearization of Eqs.~(\ref{eq_dense})--(\ref{eq_tempe}). Therefore, the turbulent structures are obtained in the context of a power balance between plasma sources, sinks (a toroidal limiter), and turbulent modes driven by the plasma gradients.

In recent years, GBS has been used to understand the nonlinear turbulent dynamics of TORPEX\cite{Fasoli2010,Ricci2008,Ricci2009,Ricci2009a} (a simple magnetized torus experiment), and the tokamak SOL in a limiter configuration\cite{Ricci2013,Halpern2013,Mosetto2013}. In particular, for both configurations the turbulent regimes\cite{Ricci2008,Ricci2009a,Mosetto2013} and the pressure decay length\cite{Ricci2009,Ricci2013,Halpern2013} have been investigated. In the present study, our simulations focus on the statistical properties of the turbulent fluctuations.

The following numerical parameters were used for the SOL turbulence simulations: $n_x=128,n_y=512,n_z=64$ ($n_x$, $n_y$, and $n_z$ are the number of radial, poloidal, and toroidal grid points, respectively). This grid results in a maximum poloidal wave number $k_{y,{\rm max}}=2$, while the largest dealiased toroidal mode number, applying the two-thirds rule, is $n_{\rm max}=21$. The physical parameters considered for the simulations are $q=4$, $\hat{s}=0$, $L_x=70$, $L_y=800$, $R=500$, $\nu=0.01$, $m_i/m_e=200$. The size of the simulation domain is equivalent to the SOL of a small tokamak such as COMPASS. \cite{Halpern2013} The source terms $S_n$ and $S_{T_e}$ in Eqs.~(\ref{eq_dense}) and~(\ref{eq_tempe}) mimic the outflow of plasma from the closed flux surface region. For simplicity, they are taken to be constant in the $y-$direction and independent of $n$ and $T_e$.

The simulations are initialized using flat smooth profiles. Then, particle and heat sources are injected, driving resistive ballooning modes linearly unstable, which in turn induces turbulent transport. As an outcome, a quasi-steady-state regime is established as a balance between the plasma sources, turbulent transport, and sheath losses. The pressure gradient length $L_p$ is not predetermined, as in linear calculations, but is instead obtained self-consistently from the calculation. In the case under study, the turbulent dynamics is dominated by resistive ballooning modes, with the non-linear saturation given by the pressure non-linearity~\cite{Ricci2013}.

\begin{figure}[h]
\centering
\includegraphics[width=0.6\textwidth]{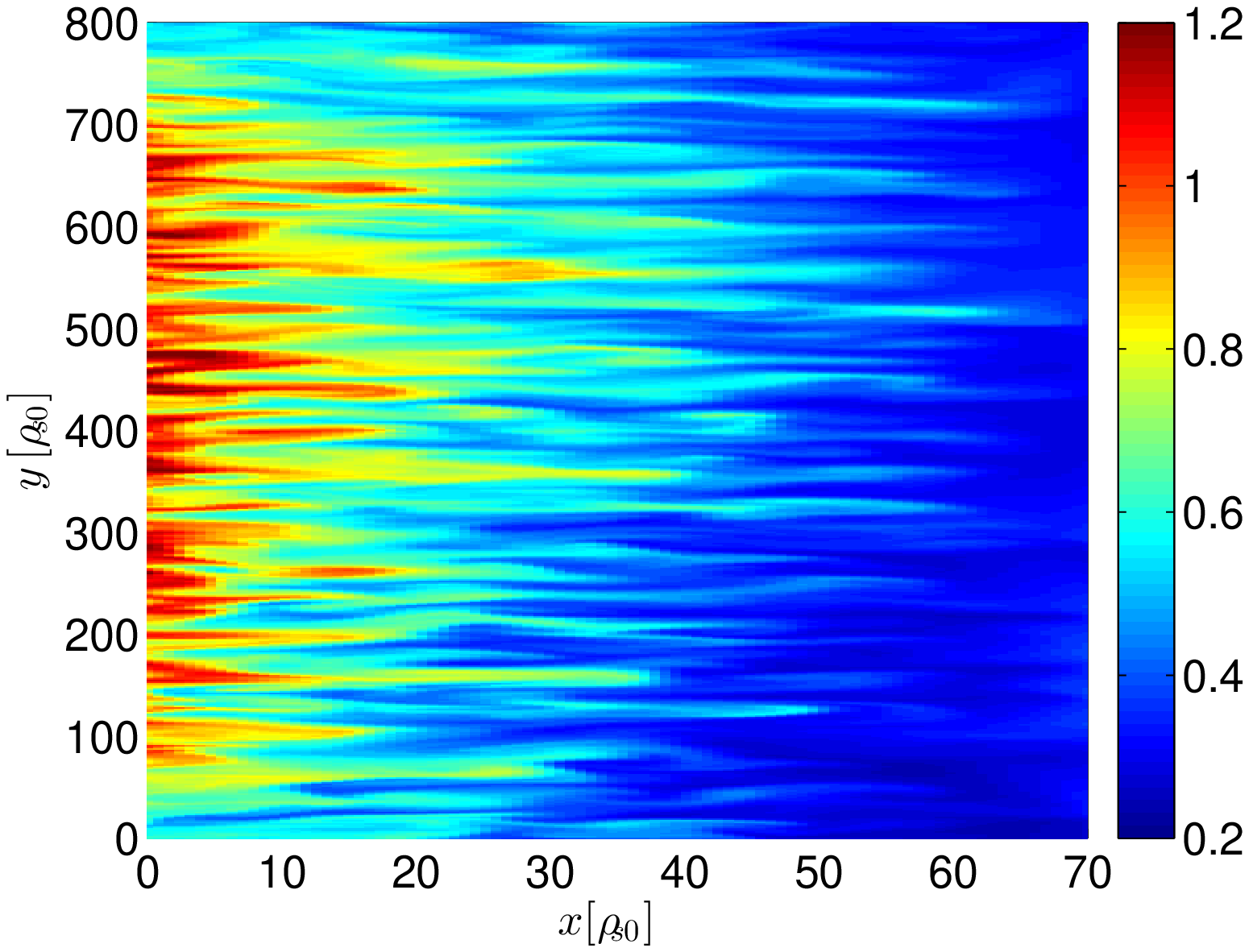}
\caption{Poloidal cross-section of plasma density as shown during the non-linear stage of GBS simulations.}\label{fig_GBS1}
\label{fig_pe_snapshot}
\end{figure}

A typical poloidal cross section of the SOL is shown in Fig.~\ref{fig_GBS1}. As it will be shown, the physics involved defines two distinct regions regarding the turbulence fluctuations in the simulation domain: At the left boundary the injected plasma is driving flute-like, radially elongated turbulent eddies, which define the near SOL. In this region, fluctuations have an essentially Gaussian PDF, and intermittent events are rare. However, as the turbulent structures propagate into the far SOL, they are sheared apart and detached blobs appear, forming the so-called 'blobby region'. Here, intermittent events become much more frequent and important as the vessel wall is approached.
In order to corroborate these statements, we have to explore the statistical properties of simulations in both the source and blobby regions, and
try to differentiate between the two. This will be achieved by means of singular spectrum analysis (SSA) \cite{ssa1}, a well-known mathematical method for
analysing the structural behavior of relatively small time-traces, by filtering out possibly existing (deterministic) oscillatory components from a weakly stochastic 
process (a typical picture is the superposition of noise on a sinusoidal signal). Such oscillatory components pertain to normal modes in the SOL simulations, which have to be removed before we embark on the statistical analysis of the GBS output data in each region of the domain. It is noteworthy, that SSA has been successfully adopted by several scientific fields, like geology \cite{ssa2}, economics \cite{ssa3} and medicine \cite{ssa4}, but hardly so in plasma physics (however, see \cite{ssa5} for a notable exception).
We shall begin our numerical study with a simulation in the source region. In Fig. 2 we show the time-trace of the density at
radial location $x=10$. We apply SSA on this signal in order to track down any oscillatory components present, which
will manifest themselves as the largest eigenvalues in the spectrum shown in Fig. 2. It turns out that the first eigenvalue is the most dominant one, and therefore the one to be removed from the time-trace. Having done this, the remaining component of the time-trace is also shown
in Fig. 2.
We will now show that the filtered data actually follows a Gaussian distribution appropriate for the weak nonlinear regime of the simplified model. For this, we employ the quantile-quantile (QQ) plot of the data
against the normal quantiles (see Fig. 4). For further ease of inspection, we superpose the Gaussian data with the same mean and standard deviation 
as the filtered data. It is clearly seen that these two sets of data almost coincide, which speaks for the Gaussian nature of the GBS filtered data.
The situation radically changes, however, in the blobby region. In Fig. 3, we present the GBS data from SOL simulations in the blobby region,
both raw and filtered, in the same fashion as before. Again, the dominant eigenvalue is the first one, which has been removed to obtain the filtered data
(see Fig. 4).
The distinct spikes in the filtered data are responsible for the emergence of a strong tail, as shown in the QQ plot (Fig. 5). In this case,
the Gaussian data with the same mean and standard deviation is clearly unable to capture the simulation data. Nevertheless, it is possible
to reproduce the tail by employing an exponential deviation to the straight line joining the 1st and 3rd quantiles of the filtered data (see Fig. 5),
where this assumption is justified by the  form of the PDF in Eq. \ref{eq:1.18}.
 
The analysis of the numerically generated data shows a distinct deviation from Gaussianity in the blob region, as seen in Fig. 5, which is a salient feature in the whole blob region whereas in the source region the filtered statistics exhibit Gaussian PDFs. This corroborates the first principles analytical modeling, suggesting that the tails of the PDFs are manifestly exponential as a result of the non-linear dynamics present in the SOL region. 
 
\section{Summary and conclusions} \label{sec4}
Transport in the tokamak scrape-off layer is dominated by intermittent and bursty processes, rendering mean-field-theory models inadequate for its 
description. In this work, we have employed the Braginskii fluid solver GBS to investigate the intermittent characteristics of the transport driven by coherent structures, such as blobs. At the same time, we derived from first principles a stochastic likelihood model of the plasma density, which is able to predict
the tails of the probability distribution function (PDF). The derivation of the model is based on the Fokker-Planck approach, yielding a closed analytical expression suitable for comparison to both numerical and experimental data. To enable such comparisons, we have processed the numerical data using the singular spectrum analysis, which filters out possibly existing oscillatory (deterministic) components from a weakly-stochastic time-trace. We have shown that the statistics of time-traces of the density can be modeled with the PDF derived from the stochastic model.
As further work, we envisage the study of higher moments, such as transport coefficients, by extending the theoretical model as well as the SSA methodology in order to handle cross-correlated time-traces.


\newpage
\begin{figure}[h]
\centering
\textbf{Raw data from SOL simulations produced by GBS in the source region and `filtered' data after the oscillatory components
have been removed. Both sets of data are normalized for zero mean and standard deviation equal to unity.} {%
  \includegraphics[width=.45\linewidth]{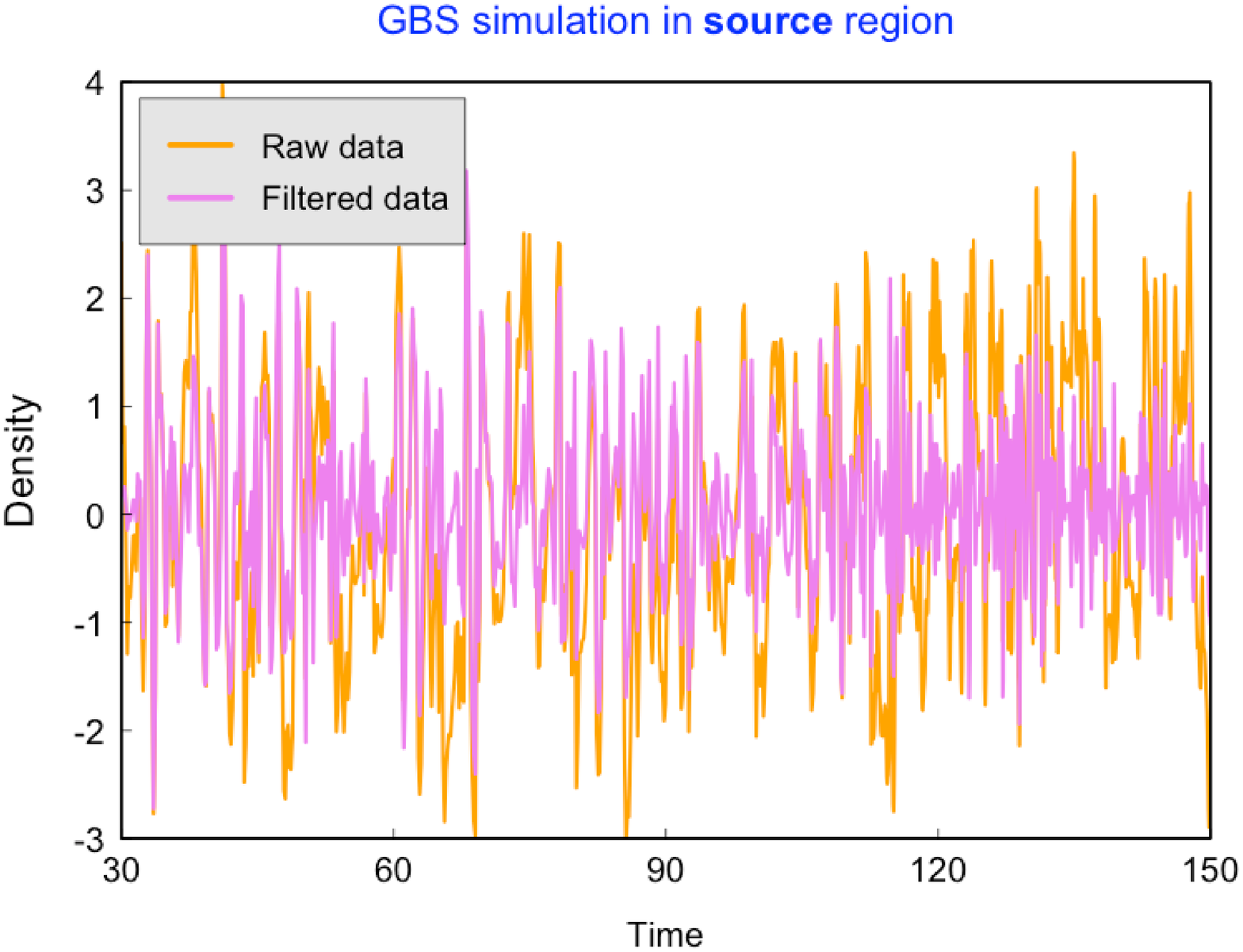}
  \label{ssaf1a}}
  \quad
\end{figure}
\begin{figure}
\centering
\textbf{Eigenvalue spectrum of the simulation data in the source region. The first eigenvalue, which is clearly the most dominant one,
has been removed from the raw data to provide the filtered time-trace.} \\ {%
  \includegraphics[width=.45\linewidth]{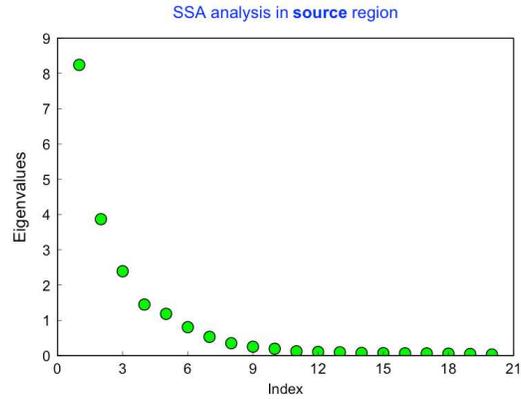}
  \label{ssaf1b}}
\caption{SSA analysis of SOL simulation in the source region.}
\end{figure}
\begin{figure}[h]
\centering
\includegraphics[width=0.6\textwidth]{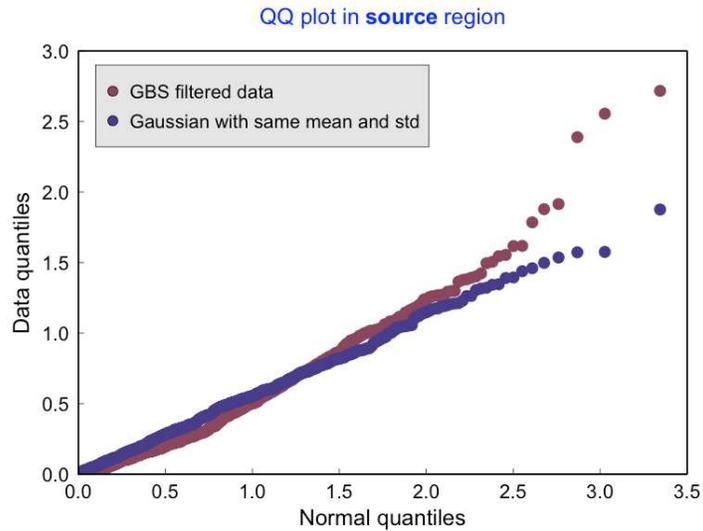}
\caption{Quantile-Quantile plot of the filtered data in the source region against the Gaussian distribution.}
\end{figure}
\begin{figure}[h]
\centering
\textbf{Raw data from SOL simulations produced by GBS in the blobby region and `filtered' data after the oscillatory components
have been removed. Both sets of data are normalized for zero mean and standard deviation equal to unity.} \\ {%
  \includegraphics[width=.45\linewidth]{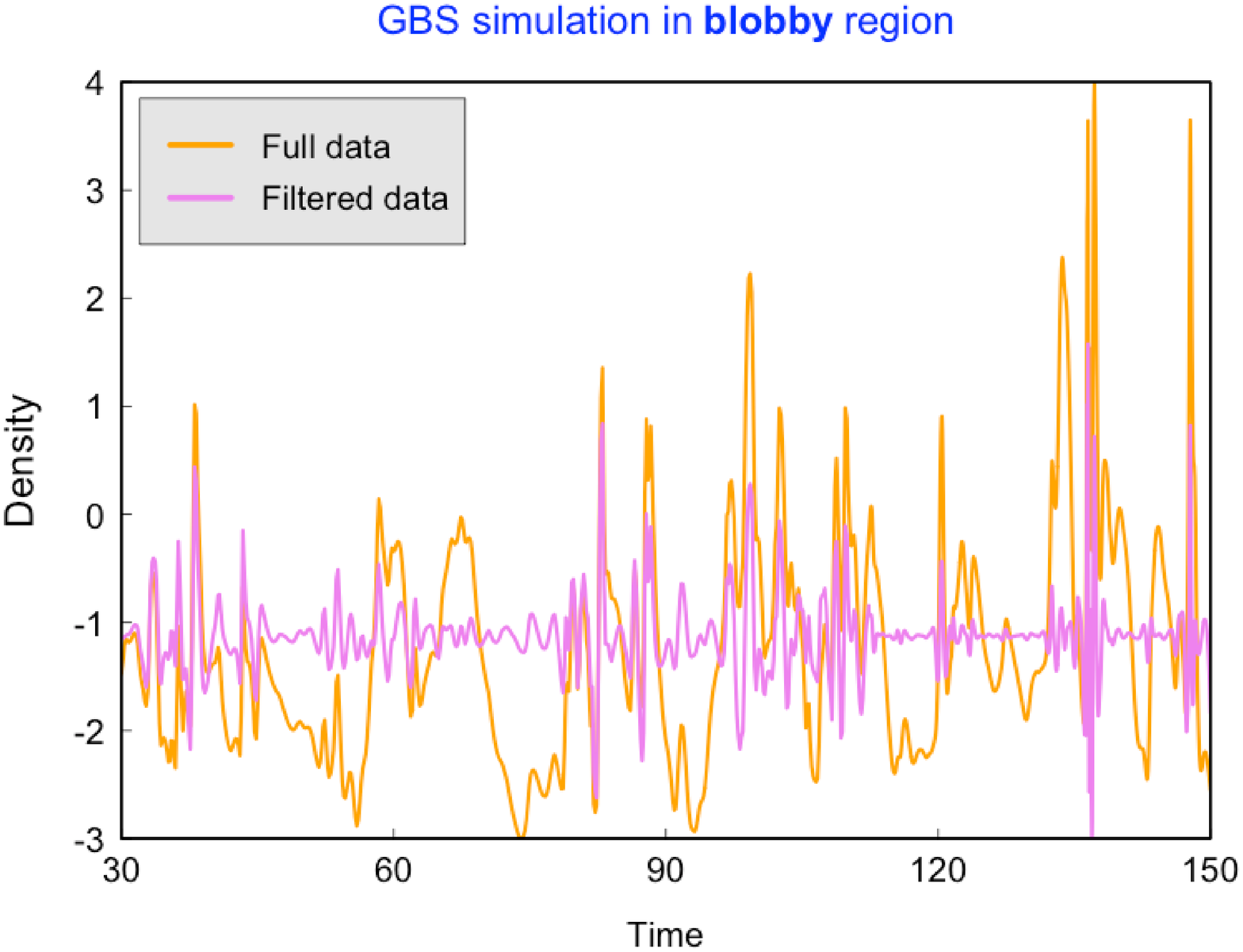}
  \label{ssaf3a}}
  \quad
\end{figure}
\begin{figure}  
\centering
\textbf{Eigenvalue spectrum of the simulation data in the blobby region. The first eigenvalue, which is clearly the most dominant one,
has been removed from the raw data to provide the filtered time-trace.} \\ {%
  \includegraphics[width=.45\linewidth]{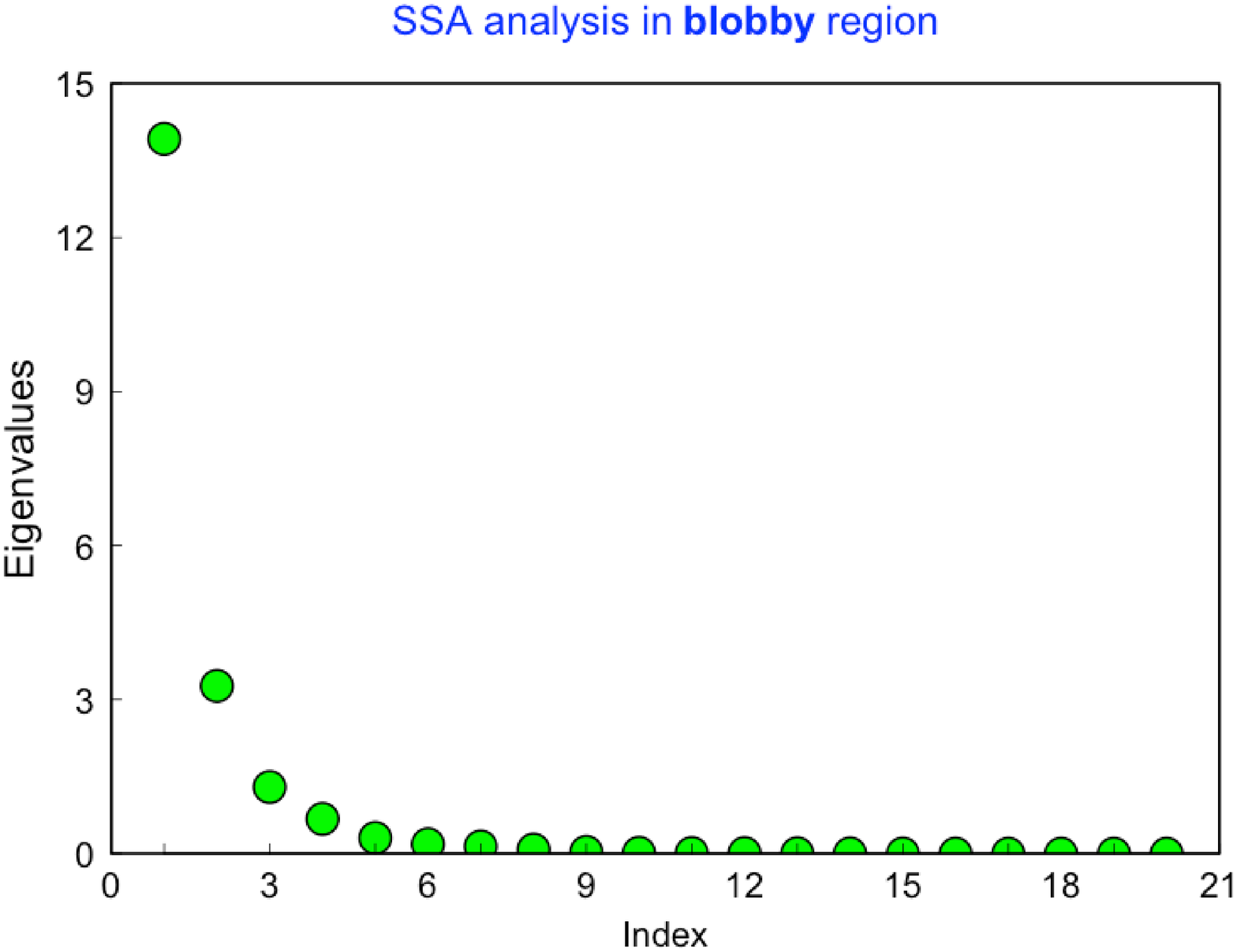}
  \label{ssaf3b}}
\caption{SSA analysis of SOL simulation in the blobby region.}
\end{figure}
\begin{figure}[h]
\centering
\textbf{Quantile-Quantile plot of the filtered data in the blobby region against the Gaussian distribution.} \\ {%
  \includegraphics[width=.45\linewidth]{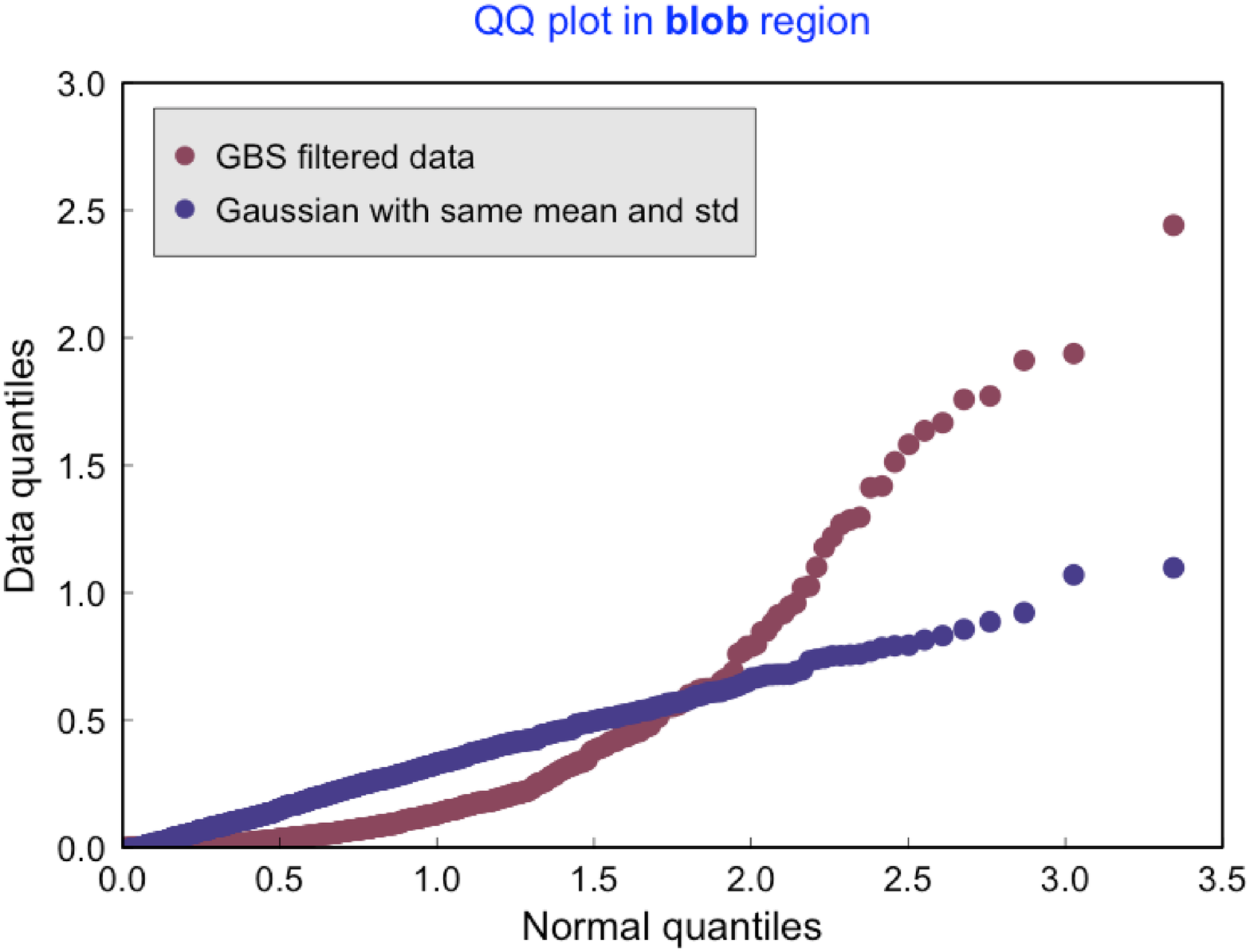}
  \label{ssaf4a}}
  \quad
\end{figure}
\begin{figure}
\centering
\textbf{Exponential fit of the tail added as a deviation to the Gaussian reference slope joining the 1st and 3rd quartiles.} \\ {%
  \includegraphics[width=.45\linewidth]{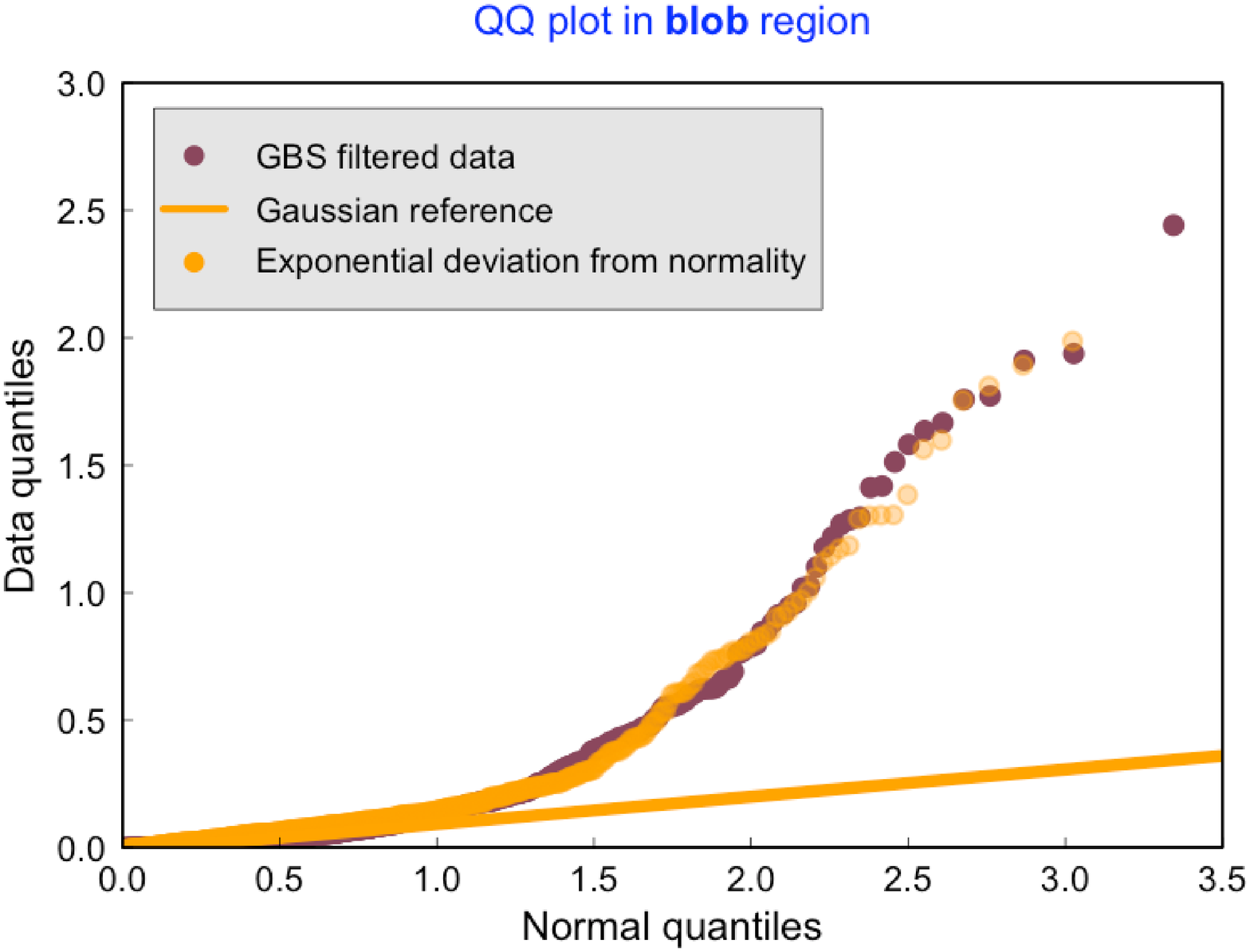}
  \label{ssaf4b}}
\caption{SSA analysis of SOL simulation in the blobby region.}
\end{figure}
\end{document}